\documentclass[10pt]{article}

\begin{document}

\title{\bf Local induction approximation in the theory of superfluid 
turbulence. Numerical consideration}
\author{M. V. Nedoboiko \\
Institute of Thermophysics, Novosibirsk, 630090, Russia }
\date{}
\maketitle

\begin{abstract}
The local induction approximation (LIA) of the Biot-Savart law is often used
for numerical and analytical investigations of vortex dynamics 
in the theory of superfluid turbulence. In this paper,
using numerical simulation, some features of the LIA is considered.
The temporal evolution of vortex loop spectrum is studied numerically.\\

\end{abstract}

\section{Introduction.}

It is well known that chaotic vortex structures,
appearing in  volume of  superfluid helium under 
the particular conditions influence on hydrodynamic and thermodynamic 
properties of HeII. At this time, any explicit macroscopic equations 
to consider
this influence are absent. In hydrodinamic approaches usually the Vinen
equation is used. The system of hydrodynamic equations for superfluid 
turbulence, where the Vinen equation is incorporated, 
was obtained by some authors. (The details can be found for example in [1][2].)
In turn, the Vinen equation is a phenomenological relation and it's validity 
is, in principle, unknown. From this point of view, for an adequate description
of various phenomena in HeII, it is desirable to consider exact microscopic
equations of vortex dynamics, based on the Biot-Savart law. But any analytical
progress in this approach is rather difficult (practically zero at this time).
Numerical investigations also have some problems. The Biot-Savart law implies the
nonlocal interaction and therefore large computer resourses are demanded.
Since the works [3][4] to the present moment, the local induction approximation
(LIA) of the Biot-Savart law is often used. But the validity of the LIA 
to describe vortex dynamics is the object of discussions until now.

This report devoted to the numerical investigation of the LIA and present the
continuation of the early analytical work [5]. 
In [5] was shown that the LIA can not describe
a flux of  amplitudes of harmonics across the spectrum in the one-dimesional
Fourier-representation. This result seems rather unusual, taking into account 
that the LIA is a nonlinear differential equation. 
The aim of the paper is to check this with help of numerical modeling.

\section{The problem statement.}
The dynamical equation of the evolution of a vortex filament has the following
LIA-form[3][4]:

\begin{equation}
\frac{d\vec{s}(\xi ,t)}{dt}=\beta \vec{s^{\prime }}\times \vec{s^{\prime
\prime }}+\nu \vec{s^{\prime \prime }}  
\end{equation}

Here $\vec{s}(\xi ,t)$ is a radius vector of a vortex line point labeled by the
variable $\xi$, $\vec{s^{\prime }}$ is the derivative on the 
parameter $\xi$; $t$ is time; the quantity $\beta $ is 
the coefficient of the nonlinearity
$\beta ~=~\frac \kappa {4\pi }\log \frac R{r_0}$, with the
circulation $\kappa $ and the cutting parameters $R$ 
(the external size, i.e. the averaged radius of
curvature) and $r_0$ (the vortex core size).
The coefficient of dissipation $\nu $ appears in the eq.(1) when 
external counterflow is absent [3][4]. 
Below one supposed that $\xi$ is the arclength.

Amplitudes of one-dimensional Fourier harmonics are defined as:
\begin{equation}
\vec{s}_k(t)\simeq \int\,d\xi\: s(t,\xi)\, e^{-ik\xi}
\end{equation}

The numerical calculations was executed for two cases. Firstly,  
the spectral evolution of some initial configuration was investigated. At
that it was supposed $\nu =0$ in the eq.(1) and thus only nonlinear term
was presented in the dynamical equation. The evolution of harmonic amplitudes
was defined by this term only. Secondly, a random force term of the Gaussian
type was added to the right hand side of eq.(1).
Thus eq.(1) was modified as:
\begin{equation}
\frac{d\vec{s}(\xi ,t)}{dt}=\beta \vec{s^{\prime }}\times \vec{s^{\prime
\prime }}+\nu \vec{s^{\prime \prime }} +\vec{f}(t,\xi) 
\end{equation}
Here $ \vec{f}(t,\xi) $ is defined by the correlator which in the 
one-dimensional $k$-space looks like: 
\begin{equation}
<\vec{f}_k(t_1)\vec{f}_{-k}(t_2)> = \frac{D}{k^y}\delta(t_1-t_2)
\end{equation}

\section{The details of the numerical procedure}
The technical detail: the figures, mentioned below are available in the individual
files. For example the file Fig.1.gif corresponds to Fig.1 etc.

In this chapter the full details of the numerical procedures are described.
A reader, not interesting this theme, may omit this section.

A vortex line was presented as a number of points in the coordinate
space $(x,y,z)$. For example, the ring with radius 1 contains 1200-5000 points,
situated on the equal distance ($\Delta s$) from each other. The fourth order 
Runge-Kutta method was used for modeling the temporal evolution. The time 
step was defined as $\Delta t = (\Delta s)^2/C$. Where $\Delta s $ is the minimal 
for the entire configuration space step, $C$ is a constant. In the work [6] one  
maintain that this method is stable when $C>\sqrt{2}$. The experimental testing
in our case gives the result $C\simeq 500-1000$ if the vortex line is strongly
intricated. The criterion of the explicity was the conservation of the full 
length
and the integral of the curvature square. (This values are some of the LIA 
invariants.) For example, this integral was equal $38222.2167961644$ and 
it was covserved with
the explicity $\pm 0.0000001$ during the program ran, the total length was
conserved down to 12 decimal signs. Besides to 
test the explicity, the iterative Crank-Nicolson type
numerical scheme was developed. This scheme was similar
to the first one, described in the work [7]. But the distinction was the 
following:
the equation for $\vec{s}$ was used, instead of the equation for a
tangent vector $\vec{l}$ as in [7]. (There method is really unusable for the 
required explicity.) The criterion of the iteration convergence was about
$10^{-18}$. Let's note that the explicity of the floating number representation
in a standard computer is about 19 characters (the format "double"). 
For example, the same initial  
configurations (it is shown at the Fig.1 on the (x,y) plane) were calculated 
with these methods till $192547$ and $1174140$ 
time steps, but for the approximately same physical time. One can see from
Fig.2 and Fig.3 (the configurations on the $(x,y)$-plane, corresponding to
these two calculations), the results are absolutely identical. The
spectral distributions are also identical. 

To create a strongly intricated initial configuration, the following 
procedure was used. Usual ring, with radius 1 was created at first.
After that, sinusoidal disturbances were added to this configuration.
The disturbances were added step by step, and after each step the procedure
of the rescaling of the length was applied, besides the coordinates of
points were corrected to place them
uniformly. In this way (step by step)  some  harmonics were excited.

When the randomized model was studied, the additional $\vec{f}(t,\xi)$ was
incorporated into the numerical method. At that this value was
modeled in the following way:
$$
 f^\alpha (t,\xi)=\sum\limits_{n=n_{low}}^{n=n_{high}}
\frac{c_t^\alpha}{(\frac{2\pi n\xi}{L_t})^\frac{y}{2}} sin(\frac{2\pi n\xi}{L_t}
+\phi _t^\alpha)  
$$
Here $(\frac{2\pi n}{L})$ is a one-dimensional wave vector $k_n$ with number
$n$, $L_t$ is a total length of a vortex line (recalculated at each time step),
$\xi$ is a current distance
along the line, to clarify $y$ see the rel.(4), $\alpha$ denotes a spatial
component, $c_t^\alpha$ is a random
amplitude, defined by  a random-number generator, 
$\phi _t^\alpha$ is a random phase. The last two values were generated at 
each time step.
 
Besides, a random disturbance
also was used to create an initial configuration. At that
a bunch of harmonics was excited, and then the random force was switched off.

During the was program running, one was possible to observe the changing of
various parameters and the spectral picture on the monitor. To
develop the program, Borland C++Builder 5 tool was used.

\section{The results, discussion and conclusions}
Thus, in the previous paper [5] was shown that the LIA don't describe a flux
of harmonic amplitudes across the spectrum. The aim of the numerical  experiment
was to confirm this result. The first part of the experiment looks as following.
One was created an initial strongly intricated configuration with
some spectral structure. After that, the vortex filament evolved accordingly
to the eq.(1) (under the condition $\nu =0$). The reconecting processes was 
omitted from the consideration,
i.e. parts of a vortex line freely passed across each other when they
intersected. The changing of the spectral picture was studed.
The initial spectral distribution ploted on the Fig.4 from $n=1$ till $n=150$
(there and below a red line
correspond to an initial state, black line denotes a current distribution).
A wave vector $k_n=2\pi n/L_t$, where $L_t$ is the total length of the filament
at the moment $t$. 
In coordinate spase this configuration plotted on the Fig.1 ((x,y)-plane). 
The successive 
steps of the spectral evolution are shown at the Fig.5-Fig.8. Finally,
the spectral distribution looks as at the Fig.9. At later time a stationary
state was formed and any changes were only inside the area, marked by blue line.
In the stationary state, the average size of the vortex tangle ceased
to grow and  only slight pulses occured.
Inside of the marked area some peaks appeared, which dassapeared
after a time. The general spectral distribution was not changed. The harmonic 
excitation
in the low region (about the first harmonic) was absent just as the excitation
in the high area. Thus, after the stationary state was formed, an 
spectrum changes was ceased.  The initial spectral distribution
was changed cardinally. One can seem that this point is not in
agreement with the
result of [5]. But the procedure, used in [5] implicity implies 
just stationary state.
As well known, a standard situation look as following. 
A wave packet spreads to infinity when a nonlinear, nondissipative equation is 
used.

The stability of the harmonic distribution, perhaps connected with the
infinite number of the invariants, produced by the LIA. For example,
total length $L=const$, $\int \, d\xi\: {\vec{s^{\prime\prime}}}^ 2=const$, etc.
High harmonics can not excitate because  it means the growth of the curvature 
and, hence, of the curvature square, this is impossible. Increasing of low
harmonics lead to the growth of a loop size, this is also impossible because
$L$ is a constant and so the total curvature must decrease in this case.
Thus some transfer across the spectrum is certanly absent 
if a vortex configuration achieved a stationary state. 

In the second part of the numerical experiment the temporary evolution 
of a detached narrow harmonic bunch was studied. The harmonics in the
environment of this bunch was equal to zero. The initial
configuration presented  a ring. Some scores of time steps this ring
undergo the random disturbances of the kind: 
$ f^\alpha (t,\xi)=\sum\nolimits_{i=n_{low}}^{i=n_{high}}
\frac{c_t^\alpha}{(\frac{2\pi n\xi}{L_t})^\frac{y}{2}} sin(\frac{2\pi n\xi}{L_t}
+\phi _t^\alpha)  $. 
The amplitudes $c_t^\alpha$
and the phases $\phi _t^\alpha$ possessed random values at each time steps. (Their
values was defined by the random number generator.) As the result
the initial spectral configuration was obtained (Fig.10). (At the Fig.10
the stripe of the spectrum is plotted from harmonic number $n=70$ up to
number $n=120$.) Later the external force was switched off and
the evolution under the LIA nonlinearity only was investigated. 

The result is following. There are no any additional harmonics exciting in
the system, on condition that they equal to zero initially.
This result was checked thoroughly. The boundaries of the excited area was
thoroughly examined after about 800 thousands time steps. They were examined at the 
enlarge scale factor  $10^{8}$ (in the vertical direction) relative to the 
amlitude of the stripe. 
As it shown at the Fig.11 where the boundaries of the excited
harmonic bunch plotted (scale factor $10^8$), there are no any disturbances
beyong the initial stripe. At the same time the spectral distribution inside
of the perturbed area was varied. This change represented the
periodical process (the period is about 20 thousands time steps). 
The steps of the changes is shown on the Fig.12-Fig.18.
Obviously, the spectral distribution strictly return to it's initial state.
The period of the return depends on the width of an excited strip and also
on current numbers of the harmonics. Note again, in a standard nonlinear
task we have a wave packet is smearing and tending to it's stationary 
spectral distribution. Note, however, after some millions time steps, distortion
of the spectral stripe was observed. Apparently it is produced not
strictly zero initial background. It was tested for the smooth ring. Obviously,
in the later case only the first harmonic must be present. But other
harmonics had amplitudes of order $10^{-7}$ (relative to the amplitudes,
plotted on the Fig.12 - Fig.18). Because, any nimerical method has
inadvertent errors, backgroud amplitudes was conserved within about 1 percent.
At the Fig.19 (scale factor $5*10^9$) shown the divergence from zero for the 
initial (red) and current (black) backgrounds. Besides, 
the graphic on the Fig.11 is wider the on the Fig.12 - Fig.18 because
the greater magnification in the vertical direction allows to see 
small amplitudes which are lost details.

Also a case when the external force not switched off was considered. The
force was during all time when the programm was running.
 At that, of course, $\nu \neq 0$
in the eq.(3). The result is the same. After the quasi-equilibrium between
pumping and dissipation was formed (i.e. when the tendency  for
changing the total length and the total curvature vanished),
any new harmonics not appeared in the system. \\

Thus two conclusions for sure following for the LIA.

1. There are no transfer across the spectrum when a vortex system is in the 
stationary state.

2. There are only initially excited harmonics remain in a vortex system. 
New harmonics never excite. All redistributions of hamronic amplitudes take 
place between excited harmonics.

And the last remark. It can be suppose that the question: "whether can LIA 
approximately describe the superflud turbulence" is rather make no sense until
we don't know exactly the role of the processes of reconnections in dynamics of
vortex loops. But this problem is too much difficult.

\bigskip
The work was carried out under
support of INTAS (grant N 2001-0618) and RFBR (grant N 03-02-16179).

\end{document}